\documentclass[12pt,manuscript,screen,nonacm]{acmart} %
\AtBeginDocument{%
  \providecommand\BibTeX{{%
    \normalfont B\kern-0.5em{\scshape i\kern-0.25em b}\kern-0.8em\TeX}}}

\copyrightyear{2025}
\acmYear{2025}
\acmDOI{XXXXXXX.XXXXXXX}

\acmConference[The 5th CI Symposium]{the 5th Annual Symposium on Applications of Contextual Integrity}{September 21--22,
  2023}{Toronto, Canada}
\acmBooktitle{5th Annual Symposium on Applications of Contextual Integrity, September 21--22,
  2023, Toronto, Canada}

\setcitestyle{numbers,square,sort&compress}

\usepackage{tikz-cd}
\usepackage{multirow} 
\usepackage{makecell}
\usepackage{amsmath}
\usepackage{enumitem}

\newtheorem{remark}{Remark}
\newtheorem{claim}{Claim}
\newtheorem{example}{Example}

\newcommand{\cut}[1]{#1}
\newcommand{\cutTwo}[1]{#1}
\renewcommand{\cutTwo}[1]{}%

\renewcommand{\epsilon}{\varepsilon}

\tikzset{
  symbol/.style={
    draw=none,
    every to/.append style={
      edge node={node [sloped, allow upside down, auto=false]{$#1$}}}
  }
}

\usepackage{lineno}

\let\small\normalsize
\let\footnotesize\normalsize

\renewcommand{\bibliofont}{\normalsize}
\renewcommand{\keywordsname}{Keywords}
\makeatletter
\let\ACM@origbaselinestretch\baselinestretch
\def\@authorfont{\normalsize\sffamily}
\makeatother

\begin{document}

\title{The Five Safes as a Privacy Context}
\author{James Bailie}
\authornote{}
\email{bailie@chalmers.se}
\affiliation{%
  \institution{Chalmers University of Technology and University of Gothenburg}
  \city{Gothenburg}
  \country{Sweden}
}

\author{Ruobin Gong}
\email{ruobin.gong@rutgers.edu}
\affiliation{%
  \institution{Rutgers University}
  \city{Piscataway}
  \state{New Jersey}
  \country{US}
}

\begin{abstract}
  \normalsize\noindent{\sffamily\bfseries Abstract}\par

  The Five Safes is a framework used by national statistical offices (NSO) for assessing and managing the disclosure risk of data sharing. It can be understood as a specialization of a broader concept---contextual integrity---to the situation of statistical dissemination by an NSO. We demonstrate this by mapping the five parameters of contextual integrity onto the five dimensions of the Five Safes. We also discuss how each of these two theories can address weaknesses in the other, thereby strengthening them both. Finally, we describe how both the Five Safes and contextual integrity can contextualize the narrow, technical notion of differential privacy, thereby guiding its implementation.
\end{abstract}


\maketitle

\section{Introduction}

As a supplier of official data, a national statistical office (NSO) is an integral part of a well-functioning democratic state. %
Its data inform government policy, business strategy and academic research, thereby supporting social progress and economic growth\cut{ \citep{lateraleconomicsValueAustralianCensus2019}}. Yet an NSO's ongoing value depends upon maintaining its social license to collect and share data. It is therefore necessary that NSOs balance their social and economic utility with the privacy of their data providers. With the recent increase in resources available to malicious actors and the growth of competing data vendors, this trade-off is increasingly difficult to manage \cite{bailieBigDataDifferential2020}.

The Five Safes (FS) is one tool that assists NSOs in navigating data-sharing decisions in the presence of potential privacy risks. 
Originally developed in 2003 to enable researchers' access to Office of National Statistics microdata \cite{desaiFiveSafesDesigning2016}, its use has since expanded to all forms of statistical dissemination \citep{australianbureauofstatisticsFiveSafesFramework2021}. It has been employed by NSOs in the UK, Australia and New Zealand to guide design decisions and risk assessments for statistical disclosure control (SDC) \cut{ \citep{stokesFiveSafesData2017, ukdataserviceWhatFiveSafes, australianbureauofstatisticsFiveSafesFramework2021, jonesBuildingAotearoaNew2022, statisticsnewzealandHowWeKeep2022}}. In the US, the FS have been used by the Coleridge Initiative for sharing data within and across states, government agencies and researchers \cite{fosterBigDataSocial2021}. Furthermore, the Advisory Committee on Data for Evidence Building recently recommended that the use of the FS in the US federal bureaucracy be expanded\cut{ \citep{advisorycommitteeondataforevidencebuildingYearReport2022}}. 

Broadly speaking, the FS is a framework 
for designing and assessing modes of statistical data sharing in ways that maintain the privacy and confidentiality of data providers. %
More specifically, it is a theory that decomposes disclosure risk into five dimensions, or `safes'---safe \emph{people}, \emph{projects}, \emph{settings}, \emph{data}, and \emph{outputs}---along with a loose set of guidelines for how an NSO can manage risk by appropriate choices in each of these five dimensions. These two parts of the FS are often conflated in existing work, %
resulting in considerable confusion surrounding what exactly the FS comprises \citep{greenPresentFutureFive2023}. 

In this work, we bring the perspective of contextual integrity (CI) to isolate and explicate the first---and most central---part of the FS: its theory of disclosure risk. We make three main points.  Firstly, FS theory may be viewed as a specialization of CI for statistical dissemination contexts. We support this claim by translating between the five CI parameters (sender, recipient, subject, information type and transmission principles) and the five `safes.' Viewed this way, the FS provides specialized guidance as to how NSOs can satisfy contextual information norms, overcoming some of the difficulties in applying CI to statistical dissemination. 

Conversely, by situating the FS within CI theory, NSOs benefit from the latter's extensive literature in justifying and understanding the FS. Our second main point, therefore, is that both the FS and CI can benefit from cross-pollination between the two fields, ultimately leading to better---i.e., safer and more useful---data access. Indeed, by stress testing our translation between CI and the FS, our analysis reveals ambiguities in the FS that are addressed by CI, and visa versa. %

Thirdly, we contextualize differential privacy (DP) by showing how it falls within the `information type' parameter of CI, as well as within the `data' and `outputs' dimension of the FS. This contextualization allows NSOs to place DP within their FS toolkit while also guiding the design of DP implementations within the broader privacy context, as delineated by both their regulation and the relevant social norms.

\cut{We conclude this article by discussing the FS as a principled yet necessarily incomplete response to the wicked problem of disclosure control, and by outlining future work on its relationship to CI, formal privacy criteria, and the legal frameworks shaping NSOs' practice.}

\section{Methods}
\subsection{Contextual integrity for statistical dissemination}\label{sec:CI}

The dissemination of statistical information by an NSO operates under a set of distinctive informational norms. In democratic societies, NSOs typically act pursuant to the authority granted by an elected governing body to administer censuses and surveys. The resulting products are disseminated in the public interest to provide a quantitative basis for evidence-based research and policy decisions. The legal, regulatory, public policy and social confines within which an NSO conducts a statistical dissemination therefore give rise to what CI describes as a \emph{context}. %

A context, according to CI, is any social setting in which information flows between parties. %
CI provides a structured way of understanding when an information flow is private based on its context.
Crudely speaking, it declares that a flow preserves privacy if the flow's context conforms to legitimate social norms \cite{nissenbaumContextualIntegrityData2019}. %
CI posits that five parameters govern such norms: the information flow's \emph{sender}, \emph{recipient}, \emph{subject(s)}, \emph{information type} and \emph{transmission principle(s)} \cite{nissenbaumPrivacyContextTechnology2010}. Importantly, whether a flow is conforms to social norms is determined by not one nor a subset of these parameters, but rather by their collective whole. 

Of the five CI parameters, the first two are the simplest to understand. The `sender' of a flow is the actor sending the information and the `recipient' is the actor receiving the information. The third parameter, the `subjects,' are those entities to whom the information pertains: often persons, sometimes businesses, and possibly even other entities, such as families or farms. The fourth parameter, the `information type,' refers to the nature of information transmitted: demographics, medical history, financial records, to name a few possibilities. It is the information's metadata---the description of how the information was collected and processed, or, in statistical terms, the information's \emph{generating mechanism}. Finally, the fifth parameter, the `transmission principles,' captures the conditions of the information flow---e.g., the permitted uses, the consent (or lack thereof) of the flow, whether the information was purchased, and so on.

The strength of CI lies in its generality and broad applicability. It has been used to explain complex privacy dynamics across a wide variety of settings \citep{nissenbaumContextualIntegrityData2019}.  However, this same generality makes it difficult to operationalize CI for some types of information flows \citep{nissenbaumOriginPrivacyProtecting, benthallIntegratingDifferentialPrivacy2024}. This has resulted in the concept of the transmission principle carrying much of the explanatory and normative burden within the framework. Yet, teasing out the transmission principles from the context of a given information flow is often challenging: rather than being formally codified, it is often implicitly hidden within the social practices, roles, relationships and tacit understandings that structure a context. Consequently, too much of CI's power hinges on how this parameter is elicited, whereas a more specialized theory---such as the FS---can provide additional scaffolding for identifying the transmission principles in specific contexts---such as those pertaining to statistical dissemination.

Another limitation of CI is that it assumes that the subjects of any information flow are always clearly delineated from `non-subjects.' Moreover, one must be able to specify the subjects of any given flow, at least in principle. These assumptions are challenged by statistical disseminations in which the transmitted information is often separated from the original data providers by a sequence of complex data processing steps (such as those found in machine learning) \citep{nissenbaumOriginPrivacyProtecting}. The problem is that  statistical information often pertains to a population of entities as a collective, rather than to a specific group of entities. As an extreme but topical example of this, consider specifying the subjects of the information contained in a large language model. %
In comparison to CI, FS theory makes no requirements that the subjects must be well specified (although in statistical disseminations where the subjects are known, they would be included as part of a FS assessment).

\subsection{Information flows in statistical dissemination and the Five Safes}
\label{sec:5s}

\begin{table*}
	\centering
    \makebox[\textwidth][c]{%
	\begin{tabular}{r|c c c c}
		\toprule  
		& \multirowcell{2}{Public use data files/\\open data} & \multirowcell{2}{User registered\\data access} & \multirowcell{2}{Physical\\data enclaves} & \multirowcell{2}{Synthetic data\\with validation} \\ \\
		\midrule
		\multirowcell{2}{Type~\ref{eq:flow_data_researcher} flow} & \multirowcell{2}{---} & \multirowcell{2}{Data are transmitted\\to registered users} & \multirowcell{2}{Microdata are accessed\\within the enclave} & \multirowcell{2}{Synthetic data are shared\\with approved users} \\ \\
		\multirowcell{2}{Type~\ref{eq:flow_output_public} flow} & \multirowcell{2}{Open data are\\disseminated online} & \multirowcell{2}{Users may publish\\their findings} & \multirowcell{2}{Summary outputs are\\exported from the enclave} & \multirowcell{2}{Validation results\\may be published} \\ \\
		\midrule \midrule
		People & NA$^*$ & Medium & High & Medium \\
		Projects & Very low & Low & High & High \\
		Settings & Very low & Low & Very high & Medium \\
		Data    & Low & High & Low & Medium \\
		Outputs & Very high & Low & High & High \\
		\bottomrule
	\end{tabular}}
	\caption{Simplified and heuristic Five Safes assessments for four different modes of statistical dissemination. \textnormal{
			To illustrate the application of the Five Safes, we rate the NSO’s degree of control over each safe across four dissemination modes from very low to very high.
			Descriptions of these modes are given in Examples~\ref{ex:open_data}-\ref{ex:synthetic_validation} respectively. $^*$The dissemination of public use data files involves a single information flow of Type~\ref{eq:flow_output_public}, hence the dimension `people'---which is a property of Type~\ref{eq:flow_data_researcher} flows---is not applicable.}}
	\label{tab:examples}
\end{table*}

Much as CI posits that five parameters govern the integrity of an information flow, the central thesis of the FS is that five dimensions of data access---\emph{people}, \emph{projects}, \emph{settings}, \emph{data}, and \emph{outputs}---collectively characterize the disclosure risk of a statistical dissemination. These five dimensions are collectively termed the five `safes' and are best reviewed by explicating the two complementary types of information flows encountered in statistical dissemination: 

\vspace{-1.5em}
\begin{align}
\text{NSO}	& \overset{\text{data}}{\xrightarrow{\hspace*{1.2cm}}} \text{people}, \label{eq:flow_data_researcher}\\
\text{NSO/people}	& \overset{\text{outputs}}{\xrightarrow{\hspace*{1.2cm}}}  \text{general public}. \label{eq:flow_output_public}
\end{align}

In a Type~\ref{eq:flow_data_researcher} flow, an NSO purveys detailed \emph{data} to a limited set of \emph{people}---e.g., a group of social scientists. Typically, these researchers must first submit an application outlining the specific \emph{project} they are working on and the data they want to analyze in this project. The NSO then determines exactly what data these people can access, and under what \emph{settings} that access is given. After the researchers access and analyze these data, they submit a second application to publish their findings. The NSO decides what \emph{outputs} can be released, which are then disseminated to the general public in a flow of Type~\ref{eq:flow_output_public}. Alternatively, Type~\ref{eq:flow_output_public} flows may occur when the data custodian directly shares information derived from their data with the general public, without involving researchers as an intermediary (as in Example~\ref{ex:open_data} below). Analogously, but less commonly, Type~\ref{eq:flow_data_researcher} flows may occur without a related flow of Type~\ref{eq:flow_output_public}, such as when an NSO makes microdata available for download to a trusted set of users (Example~\ref{ex:microdata_download}), who then do not disseminate their findings to the general public.

The FS assesses the disclosure risk of all these kinds of statistical dissemination (i.e., those disseminations that involve both types of information flows, 
as well as those that only involve a single flow of either type). %
Much like CI, the FS's parameters are related yet independent: a change in one need not affect the others. %
However, viewing these five dimensions under a joint framework allows a data custodian with a fixed amount of resources to focus on the safety of a subset of these dimensions, while maintaining control of the overall disclosure risk. The framework also allows them to weaken controls in some dimensions, at the expense of stricter controls in other dimensions---e.g., allowing access to very detailed microdata to a trusted set of people under a high degree of oversight (see Example~\ref{ex:enclaves} below).

To provide useful context for the rest of this article, we will now briefly review the meanings of each of the five `safes.'
For a given statistical dissemination, the `people' parameter of the FS measures the security of the actors receiving data in the dissemination's Type~\ref{eq:flow_data_researcher} flow. The NSO might assess security through the actors' commitment to data confidentiality and research ethics (including compliance with any stated requirements of privacy protection pertaining to the data they access), their qualifications or credentials to handle these data and their institutional backing, if any. 

The `projects' dimension evaluates whether the intended use of the data is appropriate, ethical, and compliant with relevant legislation or regulations. The use of certain sensitive data may be restricted by law to support independent scientific research only \cite[Section~12.3]{fosterBigDataSocial2021}. The data custodian would often also ascertain that their data are used toward the advancement of science, with clear and positive social benefits and in a manner consistent with modern scientific norms, including standards of reproducibility and knowledge sharing. They might also request the project's approval from relevant institutional ethics review boards. 

The `settings' parameter concerns the environment in which the information flows take place, be they physical or virtual. The key question in this dimension is, how strong are the settings in terms of their ability to restrict impermissible use, while still enabling data access for legitimate recipients? Settings include the various security measures an NSO can employ---e.g., compliance monitoring, automated auditing, firewalls, passwords and two-factor authentication, functional separation, physical or virtual data enclaves (Example~\ref{ex:enclaves}), validation servers (Example~\ref{ex:open_data}) and so on. %

The `data' dimension captures the type, sensitivity and granularity of the information transmitted in the Type~\ref{eq:flow_data_researcher} flow of the statistical dissemination. It also concerns the use of any statistical methods---such as data perturbation---to mask the personal information shared in this flow. Salient questions in this dimension include: Do the data represent a sample or a census, and if it is a sample, how was that sample selected? How much identifying information do the data contain, and how likely is it that a researcher would spontaneously recognize the subject of a data? Were the data generated by linking two or more database, or was it collected via a single instrument? When the statistical dissemination is a single flow of Type~\ref{eq:flow_output_public}, `data' describes the type and sensitivity of the microdata that generated the `outputs' (e.g., the census responses that aggregate to the disseminated population counts).

Finally, the `output' parameter of the FS is defined analogously to the `data' parameter, except applied to the information transmitted in the Type~\ref{eq:flow_output_public} flow (instead of the Type~\ref{eq:flow_data_researcher} flow). While the data in a Type~\ref{eq:flow_data_researcher} flow might consist of microdata records, the output in Type~\ref{eq:flow_output_public} flows are frequently aggregate statistics like population counts that have been protected through the addition of a controlled amount of random noise.

In principle, a statistical dissemination could be assessed against each of the five `safes' on a continuous scale. Yet, as we will discuss in Section~\ref{secDiscussion}, these dimensions are difficult to measure quantitatively; in practice an NSO instead has a list of possible choices for each of the five safes, ordered by their strength in reducing disclosure risk. The remainder of this section will provide four common examples of these different choices. These examples are summarized in Table~\ref{tab:examples}.

\begin{example}[Public use data files/Open data]\label{ex:open_data}
Statistical agencies publish public use data files for access by the general public and researchers alike. (This is an example of a flow of Type~\ref{eq:flow_output_public} without an accompanying Type~\ref{eq:flow_data_researcher} flow). Agencies do not vet people who access these files because, by design, any person or entity without abusive intentions should be able to access these public resources. In contrast, a high level of scrutiny is placed on the outputs---i.e., the information contained in this public data files---to ensure that they are not disclosive. On the other hand, since the data custodian cannot supervise the use of this information once it is made open access, no control is possible regarding the safety of the projects nor the settings in which the projects will be conducted, beyond requiring users to sign a contractual terms of use agreement.  
\end{example}

Public use data files are frequently in the form of tabular data, which are highly aggregated to ensure adequate confidentiality. Public use microdata exist too, but they are often heavily subsampled. The US Census Bureau (USCB) curates the Public Use Microdata Sample (PUMS) based on a small sample (1\% and 5\%) of responses from the American Community Survey\cut{ \citep{uscb2023acsPUMS}}. The PUMS files are available on the USCB's website and may be accessed via the file transfer protocol, a microdata analysis tool, or through an API provided by the bureau.

\begin{example}[User registered data access] \label{ex:microdata_download}
    NSOs also make some of their data products available for download by registered users (the Type~\ref{eq:flow_data_researcher} flow). These users are vetted by the NSOs, allowing for these products to be slightly more detailed compared to what could be made publicly available. However, apart from requiring registered users to agree to terms of use, NSOs have no direct control on how the users store or share these data. Additionally, the NSOs do not explicitly control the release of the outputs of each users' analyses (the Type~\ref{eq:flow_output_public} flow) in this type of dissemination, as the users are free to directly share their findings with the general public.
\end{example}

Access to the Basic Microdata products from the Australian Bureau of Statistics (ABS) are restricted to registered users. Once registered, they can download these microdata to their own computers. Because this means the data can leave the ABS's secure IT infrastructure, they highly restrict the type, granularity and detail of data available through this dissemination mode. Some of the registered users may access a Basic Microdata product, but not share the results of their analysis. In this case, there is a Type~\ref{eq:flow_data_researcher} flow without a Type~\ref{eq:flow_output_public} flow.

\begin{example}[Data enclaves]\label{ex:enclaves}
Data enclaves are secure access environments through which authorized researchers can query a custodian's database (the Type~\ref{eq:flow_data_researcher} flow). Data enclaves provide a highly secure setting for data access. Both the people and the projects seeking access are heavily vetted: only researchers who demonstrate legitimate scientific purposes of their inquiry and compliance with research ethics are allowed access. The outputs that the researcher is allowed to obtain and bring outside of the data enclave (the Type~\ref{eq:flow_output_public} flow) is subject to various degrees of scrutiny. As a result, the data accessible through enclaves can be detailed and comprehensive. 

\end{example}

Data enclaves may be physical or virtual. A physical enclave is synonymous with a research data center, such as the US Federal Statistical Research Data Center (FSRDC), or the Canadian Research Data Centre Network. 
A virtual data enclave allows authorized researchers to access restricted-use data by logging into a secure, remote server. The DataLab of the ABS is an example of a virtual data enclave.\footnote{Due to a lack of full oversight on the data access setting compared to physical data enclaves, statistical agencies debate the safety of virtual data enclaves \cite[see e.g.,][]{russell2022cdcVDE}.} 
The reader is referred to \cite{australianbureauofstatisticsFiveSafesFramework2021} for further illustrations of dissemination paradigms discussed in Examples~\ref{ex:open_data}--\ref{ex:enclaves} and an analysis of important safety considerations that pertain to them.

\begin{example}[Synthetic data with validation servers]\label{ex:synthetic_validation}
The data custodian releases a synthetic dataset that resembles the underlying confidential dataset (the Type~\ref{eq:flow_data_researcher} flow). Researchers who hold permission to access the synthetic dataset may use it to compose their desired statistical analysis including its code implementation. Then, they may \emph{validate} their results (via a ``validation server'') with the data custodian, who will run the analysis on the confidential dataset and release the outputs to the researcher (and consequently the general public) (the Type~\ref{eq:flow_output_public} flow). Before releasing these outputs, the NSO will typically apply additional controls---such as adding random noise---to ensure that they are safe.
\end{example}

A prominent case of Example~\ref{ex:synthetic_validation} is the Survey of Income and Program Participation (SIPP) Synthetic Beta (SSB)\cut{ \citep{uscb2022SSB}}. The SSB is synthesized by the USCB through integrating nine annual SIPP panels between 1984 and 2008, together with the W-2 records from the Social Security Administration and Internal Revenue Service. Researchers whose proposed analysis is deemed appropriate and feasible by the USCB are permitted to access the SSB. Prior to October 2022, access could be obtained via the Synthetic Data Server (SDS) hosted by Cornell University.\footnote{Unfortunately, the Cornell server was shut down on September 30, 2022.}  Once the researcher composes a statistical analysis program using the SSB, they submit this code to the USCB, which in turn performs the validation on the researcher's behalf on the Gold Standard File (GSF)---the confidential microdata. The output is subject to a stringent level of disclosure review similar to those applicable to the FSRDCs.

Example~\ref{ex:synthetic_validation} is an interesting mode of statistical dissemination, through which we can observe a juxtaposition of safety levels pertaining to the two information flows.  In the Type~\ref{eq:flow_output_public} flow, the receivers are the general public and the outputs are strictly scrutinized according to a high safety standard, rendering this setting similar to the open data mode discussed in Example~\ref{ex:open_data}. In the Type~\ref{eq:flow_data_researcher} flow, the transmitted data consists of the SSB, which commands a level of safety higher than simply sharing the GSF due to its synthetic nature. The people, here referring to the researchers who access the SSB, are placed under a moderate level of scrutiny. The setting, the Cornell SDS, is effectively a virtual enclave. These elements render this setting analogous to the data enclave mode discussed in Example~\ref{ex:enclaves}. 

\subsection{The Five Safes as a privacy context for statistical dissemination}\label{sec5SafesCI}

The previous section argued two points: First, a statistical dissemination by an NSO consists of a Type~\ref{eq:flow_data_researcher} flow followed by a related Type~\ref{eq:flow_output_public} flow, or a single flow of either type. Second, FS posits that the five `safes' are the parameters that govern the disclosure risk of these two flow types. This naturally leads to the following claim, which embeds %
FS theory into the broader framework of CI: %

\begin{claim}\label{claim:5S-specialization}
    The FS is a (faithful, albeit inexact) specialization of CI to the dual information flows of statistical dissemination.
\end{claim}

To unpack this claim, we map the meanings of the five CI parameters in statistical dissemination contexts onto the five `safes.' This map is summarized in Table~\ref{tab:CI_5S}. We characterize the FS as a `faithful albeit inexact' specialization because while there some ambiguities in this map, the general character of CI is maintained. Although an attentive reader will no doubt pick up on these ambiguities in what follows, we postpone their discussion to Section~\ref{sec5SAmbiguities} where we stress test Claim~\ref{claim:5S-specialization}. %

\begin{table*}
	\centering
	\begin{tabular}{r|l}
		\toprule  
		Informational norm parameters & Their meanings in statistical dissemination \\
		\midrule
		sender    &  NSOs~\eqref{eq:flow_data_researcher} and {\bf people} or NSOs~\eqref{eq:flow_output_public} \\
		recipient &  {\bf people}~\eqref{eq:flow_data_researcher} and general public~\eqref{eq:flow_output_public} \\
		subject(s)  &  is a component of {\bf data}~\eqref{eq:flow_data_researcher} \\
		information type & is a component of {\bf data}~\eqref{eq:flow_data_researcher} and {\bf outputs}~\eqref{eq:flow_output_public} \\
		transmission principles & encompasses {\bf projects}, {\bf settings}, and  more \\
		\bottomrule
	\end{tabular}
	\caption{The parameters of contextual informational norms and their meanings in statistical dissemination, \textnormal{with reference to the Five Safes (in bold) and relevant information flows (Type~\ref{eq:flow_data_researcher} or~\ref{eq:flow_output_public}). We use the term NSOs as shorthand for statistical agencies and data custodians more generally.}
	}
	\label{tab:CI_5S}
\end{table*}

Of the five CI parameters, the first two
are most straightforwardly mapped to the five `safes.' %
The `sender' in a Type~\ref{eq:flow_data_researcher} flow is the NSO, or more generally, the statistical agency or custodian of the data being disseminated. In Type~\ref{eq:flow_output_public} flows, the `sender' is either this same NSO, or the `people' who received the `data' in the corresponding Type~\ref{eq:flow_data_researcher} flow. That there are two options here reflects the fact that there are two possible scenarios: either the researchers (i.e., `people') share their findings directly with the general public (as in Example~\ref{ex:microdata_download}), or their outputs are released by the NSO (as in Examples~\ref{ex:enclaves} and~\ref{ex:synthetic_validation}). The `recipients' of a statistical dissemination are the `people,' for the Type~\ref{eq:flow_data_researcher} flow, and the general public, for the Type~\ref{eq:flow_output_public} flow. 

The latter three parameters of CI require more explication. On face value, the two information flows of a statistical dissemination share the same `subjects,' since they are both based on the same underlying data. Nevertheless, as the `data' shared in a Type~\ref{eq:flow_data_researcher} flow is often census or survey questionnaire responses, it is usually easier to understand who its subjects are by associating each response with the entities
it measures. 
In comparison, it can be difficult to determine the `subject(s)' of a Type~\ref{eq:flow_output_public} flow, as the `outputs' are often the result of a statistical analysis concerning a larger underlying population that is represented by the subjects of the corresponding Type~\ref{eq:flow_data_researcher} flow.
For these reasons, the concept of `subjects' is often not appropriate for Type~\ref{eq:flow_output_public} flows \citep{nissenbaumOriginPrivacyProtecting}, but usually falls within the `data' dimension for Type~\ref{eq:flow_data_researcher} flows. %
That said, the FS does not require the subjects of a Type~\ref{eq:flow_data_researcher} flow to be specified, and there are cases where such a requirement would be nonsensical---e.g., when releasing synthetic data (Example~\ref{ex:synthetic_validation}).

Because the two flows of a statistical dissemination transmit different kinds of information, they each have their own `information type.' 
The `information type' of a Type~\ref{eq:flow_data_researcher} flow is a component of `data' since the nature of the information transmitted concerns both the structure and the substance of the `data.' On the other hand, the `information type' of a Type~\ref{eq:flow_output_public} flow is a component of `outputs' because the custodian's decision about what to release to the general public includes a determination on its information type.

For a statistical dissemination, the transmission principles encompasses two dimensions of the FS: `projects' and `settings.'  As described in Section~\ref{sec:5s}, to assess whether a project is safe is to ask whether the dissemination is used for an appropriate purpose---be it to support scientific research, serve the public interest, or enable evidence-based policy making---and to ensure compliance with legal and regulatory mandates. The intended use of the dissemination delineates the scope of the project, and in this capacity, the purpose partially determines the transmission principles. Similarly, `safe settings' are concerned with the environment of the dissemination---e.g., whether it is conducted in a data enclave. Clearly, these settings are also conditions on how the information flows from NSO to the `people' or general public and as such, would be classified under CI as transmission principles.

Until now, we have followed the standard presentation of the FS as a risk assessment tool. In contrast, CI is a normative theory---it is concerned with explaining and describing privacy in terms of social norms. While practitioners of the FS would typically ask questions of the form, ``Are the values of the five `safes' sufficient to ensure that the risk of a privacy violation is acceptably low?'', the standard question in CI is, ``Are the values of the CI parameters appropriate in the sense that they conform to ethically legitimate contextual informational norms?''

On first glance, these differing perspectives pose a challenge to our claim that the FS is a specialization of CI (Claim~\ref{claim:5S-specialization}). Yet, in the case of statistical dissemination, these perspectives are two sides of the same coin. The ethical responsibility of an NSO is to mitigate disclosure risk while simultaneously supplying access to its data in the promotion of the public good. As such, an NSO's disclosure risk assessment is precisely a question of whether it is meeting its normative duties. This reasoning is the basis for the following claim:

\begin{claim}\label{claim:5S-norms}
        The five `safes' %
        can be viewed as the parameters of the contextual informational norms that define the appropriateness of a statistical dissemination. %
\end{claim}

Because it reorients the FS as a normative theory, Claim~\ref{claim:5S-norms} underpins our view of the FS as a specialization of CI (Claim~\ref{claim:5S-specialization}). It is supported by the idea that the FS is not simply a risk assessment tool, but also a framework for trading off the utility of an NSO's data assets with the consequent disclosure risks. For example, it is well recognized in the FS literature that a project that is of great public interest can offset riskier choices for the other four `safes.' This kind of trading off of competing interests is not valid from the perspective of risk assessment (as public interest clearly does not reduce risk); rather, it is an example of the calculus used in CI for evaluating the legitimacy of privacy norms \citep{nissenbaumContextualIntegrityData2019}.

\section{Results}
\subsection{What can contextual integrity do for the Five Safes?}\label{secFiveSafesLearn}

\subsubsection{Grounding the Five Safes in normative theory}

The FS is primarily described as a risk assessment tool. As we have argued (see Claim~\ref{claim:5S-norms}), however, it may equally be viewed as a theory for describing the privacy norms of a statistical dissemination. Recasting the framework in this way allows the FS to draw on insights from CI literature, which offers tools for understanding and evaluating the appropriateness of an information flow (see Section~\ref{subsecElicitChoices} below). Furthermore, CI brings a normative dimension that can legitimize an NSO's choices for the five `safes' by anchoring them in shared expectations about acceptable dissemination practices, rather than in abstract notions of risk alone.

Grounding the FS in a normative theory helps elucidate the best practices with its application. The FS framework has been criticized because it is unclear how risks from across its five dimensions should be aggregated into a holistic measure. Indeed, Culnane et al. \cite{culnaneNotFitPurpose2020} argue that a single vulnerability in any one of the five `safes' can render the dissemination unsafe. Regardless of the validity of this argument, framing the FS as a normative theory offers a way to sidestep this difficulty. Rather than attempting to compute an overall measure of risk, the framework can be seen as articulating conditions under which information flows are socially, legally, or normatively appropriate. In this light, it could be more productive to treat the FS not as a model of risk, but as a normative framework of appropriate statistical dissemination, based on existing theory developed by CI.

\subsubsection{Eliciting appropriate choices for the five `safes'}\label{subsecElicitChoices}

The integration of CI’s empirical toolkit with the FS offers a principled means of aligning statistical disclosure practices with societal expectations. CI has been used extensively to elicit and measure privacy norms and expectations, both through quantitative approaches such as vignette-based surveys \citep{martinDiminishedJustDifferent2012} and through qualitative methods such as interviews and focus groups \citep{kumarRoadmapApplyingContextual2024}. These methods enable researchers to evaluate the appropriateness of specific information flows.

By considering the FS as a normative theory, NSOs can apply these techniques to elicit socially grounded choices for each of the five `safes.' Rather than relying solely on expert judgment or procedural tradition, decisions about data access, researcher vetting, project legitimacy, data treatment, and dissemination channels could be informed by evidence of publicly endorsed norms of appropriateness.

\subsubsection{Framing statistical dissemination as information flows}
CI's concept of an information flow is useful for describing the FS because it makes explicit the dual flows of statistical dissemination. Expressing statistical dissemination in this way makes it clear that the FS concerns two separate, yet related, flows. Building the framework of the FS through these dual flows facilitates risk assessments in much the same way that the decomposition of risk into five `safes' does.

Furthermore, framing the FS in terms of information flows separates its theoretical core from its practical application. In practice, the FS is operationalized through guidelines---heuristics and principles for assessing risk in each of the five `safes.' Most criticisms of the FS target these guidelines, without recognizing that it is the theory of the FS that is the more fundamental and, we contend, more useful part of the framework \citep{culnaneNotFitPurpose2020, coleDesigningAccessDifferential2020}. By clearly separating theory from guidelines, we hope to clarify critiques of the FS and to refocus attention away from debates over the validity of particular guidelines, thereby enabling improvements to both its theoretical framework and its practical guidance.

\subsubsection{Uncovering ambiguities in the Five Safes}\label{sec5SAmbiguities}

By analyzing the FS from the perspective of CI, we identify three ambiguities in its theory---that is, three 
aspects of an NSO's statistical dissemination that are not considered explicitly by the FS. The identification of these ambiguities clarifies the scope of the FS and in so doing helps strengthen its theory. It also illustrates the fundamental limitations in mapping the FS onto CI (i.e., it qualifies Claim~\ref{claim:5S-specialization}).

Firstly, a FS assessment does not explicitly specify the sender of a Type~\ref{eq:flow_output_public} flow. In some situations, this sender will be the NSO---e.g., when the NSO releases public use data files (Example~\ref{ex:open_data}). In other situations---e.g., when trusted users can download and analyze microdata (Example~\ref{ex:microdata_download})---the sender will be equal to the `people' parameter of the FS assessment. And finally, the sender of a Type~\ref{eq:flow_output_public} flow could alternatively be both the NSO and the `people' acting together. This can occur when the `people' is working inside a data enclave (Example~\ref{ex:enclaves}), as they typically must negotiate with the NSO about what outputs will be released (e.g., the NSO may give the researcher a `budget,' which they can `spend' on releasing more outputs of lower accuracy, or fewer outputs of higher accuracy). Although not captured by the FS, clearly, from the perspective of CI, it matters which out of the three options the sender of a Type~\ref{eq:flow_output_public} flow is.

Secondly, CI recognizes that there are many different ways to `release outputs to the general public'---and indeed, many different ways to interpret what the `general public' means. There are evident differences between publishing an output statistic on the front page of the New York Times as compared to burying it at the end of a hundred page policy document, even if the general public can access the policy document. How outputs are disseminated to the public matters. This nuance is missed by the FS because it does not distinguish between the different ways of publishing statistics. Moreover, it is not always the case that `people' will disseminate their findings to the general public. For example, a registered user (Example~\ref{ex:microdata_download}) could share their outputs only with a business interest group. This changes the `receiver' of the Type~\ref{eq:flow_output_public} flow from the general public to this interest group. Therefore, should FS theory make explicit two sets of `receivers'---one for the Type~\ref{eq:flow_data_researcher} flow and one for the Type~\ref{eq:flow_output_public} flow?

Thirdly, it is important to recognize that the FS is a \emph{self}-assessment. 
The `sender' of the Type~\ref{eq:flow_data_researcher} flow is not explicitly stated, as it is taken as given that the NSO undertaking the assessment is the entity disseminating the data. Consequently, the `sender' is not treated as a parameter within the FS, while it is in CI. There are also parts of statistical dissemination's transmission principles that are not treated by the FS because they are assumed to be fixed or premised by the principles and objectives of the NSO. Moreover, the operating environment of the NSO---e.g., its governmental or legislative legitimacy and its social license, or lack thereof---is not explicitly addressed in the FS.

The idea that the FS should always be viewed as a self-assessment is complicated by the fact that NSOs frequently acquire administrative data from other government agencies (e.g., as part of the SSB outlined after Example~\ref{ex:synthetic_validation}). 
In such cases, the NSO could be understood as an intermediary rather than the sender. In fact, the transfer of administrative records to the NSO should be viewed as a distinct information flow---a flow which is relevant for assessing the disclosure risk of the statistical dissemination, but which is not captured within the FS framework.

\cutTwo{The transmission principles of a statistical dissemination may also entail other aspects that are not easily captured within the FS. An example may be the distinct authority by which the NSO disseminates certain data. Take, for example, the USCB's constitutional mandate to enumerate the population every ten years for the purpose of apportionment. %
For this reason, the public dissemination of state population totals is exempt from any SDC protection, even though exact statistics may carry increased risk of disclosure. The FS framework may have to acknowledge the dissemination of state population totals as a case in its own right, but the apparently risky dissemination is entirely appropriate when viewed under the pertinent transmission principle, namely, the bureau's charge to enable a constitutionally sanctioned political process.}

\cut{More generally, the theory of the FS could be enriched by leveraging the extensive discussion in CI on the different kinds of transmission principles. This may lead to additional `safes' that more explicitly describe the transmission principles of a statistical dissemination (as proposed in \cite{oppermannPrivacyPreservingDataSharing2019}).}

\subsection{What can contextual integrity learn from the Five Safes?}\label{secCILearn}

\subsubsection{Connecting the dual information flows of statistical dissemination}

As we have described, a statistical dissemination is 
often comprised of two sequentially dependent information flows: a Type~\ref{eq:flow_data_researcher} flow, in which data are transmitted to accredited researchers under controlled access arrangements, and a Type~\ref{eq:flow_output_public} flow, in which derived findings are disseminated to the public. These two flows form a causally and normatively linked system: the legitimacy of the public release depends on the propriety of the underlying access, and conversely, the justification for researcher access is often premised on the social value of eventual publication.

CI offers limited traction for analyzing such multi-stage or coupled information flows. It treats each flow as analytically discrete, even when flows are operationally and ethically entangled. The FS framework, on the other hand, is explicitly designed to manage the interdependence of a dissemination's dual flows by modeling them as a coordinated process in which safeguards and social value are distributed across both flows.

Another, related benefit of the FS lies in its treatment of an information flow's `subject(s).'  As discussed in Section~\ref{sec:CI}, a CI analysis of a flow presupposes the existence of discernible `subjects.' This assumption is often violated by Type~\ref{eq:flow_output_public} flows, where the connection to individual data providers is oblique or diffuse (see Section~\ref{sec5SafesCI}). The FS offers a pragmatic solution by linking the two flows explicitly: the subjects of a Type~\ref{eq:flow_data_researcher} flow are indirectly implicated in the corresponding Type~\ref{eq:flow_output_public} flow, thereby resolving the difficulty often encountered in specifying the `subjects' of a Type~\ref{eq:flow_output_public} flow.

\subsubsection{Explicating transmission principles in statistical dissemination}

In CI theory, the `transmission principle(s)' parameter bears a substantial analytical burden\cut{, to the point that some scholars have proposed augmentations and extensions to CI \citep{benthallContextualIntegrityLens2017, benthallIntegratingDifferentialPrivacy2024} to explicate aspects of an information flow that might otherwise by understood as transmission principles}. 
This makes CI less tractable in specific scenarios, like statistical dissemination, where the relevant flows are structured by well-defined institutional practices and procedural safeguards. 

In comparison, the FS framework offers a domain-specific instantiation of CI that makes the abstract elements of a statistical dissemination's transmission principles more explicit and actionable. Specifically, the `projects' parameter foregrounds (among other things) how the dissemination is to be used. This is a critical determinant of the ethical legitimacy of an information flow \citep{nissenbaumContextualIntegrityData2019}, which is left implicit in CI’s abstract formulation of transmission principles. Moreover, the `settings' parameter explicates the IT security controls put in place by the NSO to protect the dissemination. These two parameters detail types of transmission principles, enabling more fine-grained reasoning about the appropriateness of a statistical dissemination and its associated disclosure risk. 
However, as noted in Section~\ref{sec5SAmbiguities}, `settings' and `projects' together do not exhaust the full scope of a statistical dissemination's transmission principles. They provide a structured decomposition relevant to the dissemination context, but they are not a substitute for the broader viewpoint provided by the concept of the transmission principle. \cut{This discussion illustrates our larger point that both frameworks---CI and the FS---have much to gain through the cross-fertilization of their separate bodies of literature, a subject that future work could further develop beyond the limited analysis given here.}

\subsection{Analyzing differential privacy with the Five Safes and contextual integrity}\label{sec:DP}

DP is a state-of-the-art family of mathematical formalizations of data privacy \citep{dwork2014algorithmic}. It has been adopted by technology companies and NSOs, most notably for the 2020 US Decennial Census. Since the proposal of $\epsilon$-DP (or pure DP) \citep{dwork2006calibrating}, over 200 versions of DP have been invented \citep{desfontainesSoKDifferentialPrivacies2022}, including probabilistic DP \citep{dworkOurDataOurselves2006}, approximate DP \citep{machanavajjhala2008privacy}, zero-concentrated DP \citep{bunConcentratedDifferentialPrivacy2016a}, and $f$-DP \citep{dong2022gaussian}, to name but a few. While the exact mathematical definitions of all these versions differ, their essence remain the same: Every one of them is a technical standard that requires the rate of change---or `derivative' (hence the epithet `differential')---of an \emph{SDC mechanism} to be controlled below a specified level \citep{bailieRefreshmentStirredNot2025c, bailieRefreshmentStirredNot2025b}. 

An SDC mechanism is any algorithm that takes a dataset as input, transforms that dataset with the goal of reducing disclosure risk, and then outputs the transformed data. 
(An SDC mechanism is also called a data release mechanism, although this is inaccurate since a data release mechanism does not actually release data but instead prepares data for release.)
The 200+ definitions of DP differ in how they measure a mechanism’s derivative, with each definition providing a different criterion against which an SDC mechanism can be assessed.
Yet, with any of these definitions, an SDC mechanism either satisfies that DP definition---because its derivative is below the level set by that DP definition---or it does not satisfy that definition---because its derivative goes above the maximum allowed level.  
Accordingly, when a statistical dissemination---such as the 2020 US Census redistricting data---is described as ``DP,'' what is meant is that the disseminated data are the output of some SDC mechanism that satisfies some DP definition.

Importantly, SDC mechanisms are applied \emph{before} any data are released, shared, or otherwise transmitted.
That is to say, an SDC mechanism is not a part of an information flow, but rather a transformation that precedes it.
For example, a public use data file (Example~\ref{ex:open_data}) is usually created by applying an SDC mechanism to confidential data, with the resulting file then made available to the public. It is this latter action---the public release of the file---that constitutes the information flow, not the preceding mechanism. %

Scholars have attempted to incorporate DP into CI by proposing a new parameter, such as the `transmission properties' of a flow \citep{benthallIntegratingDifferentialPrivacy2024}. Alternatively, one might consider DP as potentially having some bearing on a flow's transmission principles. Our view, on the contrary, is that DP is not a condition on---nor a property of---the information flow. Rather, DP is a property of the transmitted information:\footnote{Note that, when speaking about DP, we are restricting our attention to the two standard trust models of DP (central and local DP). While all implementations of DP currently listed in the DP Deployments Registry \citep{DifferentialPrivacyDeployments} and in the majority of DP literature use one of these standard models, there are other trust models---specifically, the shuffle, distributed and pan-privacy models---for which our discussion does not hold. The shuffle and distributed trust models of DP impose conditions on the transmission itself (e.g., that the flow goes through a trusted intermediary); and the pan-privacy model imposes conditions on computations downstream of the transmission.} 
\begin{claim}\label{claim:DP-CI}
    DP falls entirely within the scope of CI's `information type' parameter.%
\end{claim}
In order to substantiate this claim, recall a point we made earlier: A DP designation signifies that the information being transmitted was generated by a SDC mechanism that satisfies some DP definition. This SDC mechanism---and consequently, any DP designation---is therefore a component of how the transmitted information was created.
Since the `information type' parameter includes, by definition, the metadata about how the transmitted information was generated, an information flow ``being'' DP is in fact a statement about the flow's information type.
That is to say, being a partial description of how the transmitted information was produced, DP is simply a property of that information and, as such, is wholly contained in the `information type' parameter.

While Claim~\ref{claim:DP-CI} situates DP within the framework of CI, it is still informative to expand on the roles DP can and cannot play in statistical dissemination through the lens of a more-specific theory. Indeed, the FS delineates a context in which DP can be understood. What we mean by this is the following:

\begin{claim} 
    DP should be viewed as a quantitative standard of safety pertaining to certain aspects of the \emph{outputs} and \emph{data} `safes.' 
\end{claim}

A quantified measure of the strength of parts of the outputs and data is helpful, because it allows for better modulation of the five `safe.' %
Such modulation may be achieved in multiple ways, two of which we discuss here. 

First, DP acts as a ``screen'' between the NSO's underlying, confidential data asset and the individuals who access that data. %
By construction, a DP definition constrains the SDC mechanism which is used to create the `data' or `outputs.' %
As such, DP may be employed by an NSO to create public use data files (as we noted above) as well as to restrict the information released in researchers' public findings after they access data enclaves or validation servers \citep{bailieABSPerturbationMethodology2019}. DP may also directly restrict how researchers (i.e., `people') may interact with the data in the first place---e.g., by giving them access only to differentially private synthetic data rather than the real data \cite[see e.g.][]{bowen2022synthetic}. %
In each of these cases, a DP designation provides a mathematically rigorous characterization of the difficulty adversarial agents face in deducing the value of the confidential data supplied as input to the SDC mechanism.
Such strong quantitative controls of the `data' and `outputs' dimensions enable, at least in a heuristic way, less scrutiny to be placed on the other three dimensions of the FS---`people,' `projects,' and `settings.'

Second, DP enables quantitative ``privacy accounting'' by NSOs. This means two DP definitions of the same flavor, associated with two separate data releases, may be combined to yield an overall DP guarantee which assesses the disclosure risk pertaining to both releases jointly. %
What this means is that the data custodian with a fixed risk budget may choose to divide that budget across a number of \emph{projects}, evenly or unevenly according to their significance, modulating the quantity--quality tradeoff in ways that they see fit.

We also observe two limitations of DP as a quantitative standard of disclosure risk.

\begin{remark}
DP is silent on certain aspects of CI's information type, and of the outputs and data `safes.'
\end{remark}

DP partially characterizes the protection afforded by an SDC mechanism.
It is, however, agnostic to the \emph{nature} of the dataset input into that mechanism. %
That dataset may be highly sensitive in nature (e.g., records of patients suffering from a socially stigmatized disease) or completely innocuous (e.g., a log of customers' dairy preferences at a coffee shop).
In either case, however, the resulting data release will carry the same DP standard, provided that it is generated by the same SDC mechanism---or, indeed, by any other mechanism satisfying that standard.
Therefore, DP should not be taken as a comprehensive quantification for the safety of the outputs and the data.

\begin{remark}
DP does not purport an assessment of safety for people, projects, or settings.
\end{remark}

As we have repeatedly emphasized, DP is solely a property of the process through which the data or outputs were generated, with no consideration to the broader environment in which this generation happened. As such, it is agnostic to the settings and projects in which the statistical dissemination takes place. %
It is also agnostic to the people receiving the data. 

Our aim with this section is to provide an initial, high-level contextualization of the narrow notion of DP within the holistic theories of FS and CI. (The relationships between these concepts are summarized in Figure~\ref{fig:DP}.) 
Although the connections between these concepts warrant further, in-depth examination, the potential payoff of such a contextualization is twofold.
First, it will clarify the extent to which DP can mitigate risks arising from the other safes, as well as risks associated with aspects of data and outputs that DP does not directly address. Second, it will determine how the broader context---including the other safes and the non-DP aspects of `data' and `outputs'---should inform the choice of DP definition.

\begin{figure*}
\centering
\begin{tikzcd}[row sep=1cm, column sep=1cm,every cell/.append style={align=center}]%
    \begin{tabular}{c}
        Differential \\ Privacy
    \end{tabular} 
    \arrow[r, hookrightarrow] \arrow[d, Mapsto] &

    \begin{tabular}{c}
        Safe Data / \\ Safe Outputs
    \end{tabular} 
    \arrow[r, hookrightarrow] \arrow[d, Mapsto] & 
    
    \begin{tabular}{c}
        The Five \\ Safes
    \end{tabular}
    \arrow[r, hookrightarrow] \arrow[d, Mapsto] & 
    
    \begin{tabular}{c}
        Contextual \\ Integrity
    \end{tabular} 
    \arrow[d, Mapsto]
    
    \\
    
    \begin{tabular}{c}
        SDC \\ Mechanism %
    \end{tabular} 
    \arrow[r, hookrightarrow] &
    
    \begin{tabular}{c}
        Information \\ Type
    \end{tabular} 
    \arrow[r, hookrightarrow] &
    
    \begin{tabular}{c}
        Type~\ref{eq:flow_data_researcher} and/or \\ Type~\ref{eq:flow_output_public} Flows
    \end{tabular} 
    \arrow[r, hookrightarrow] &
    
    \begin{tabular}{c}
        Information \\ Flow
    \end{tabular}
\end{tikzcd}
    \captionsetup{singlelinecheck=false}
    \caption[.]{%
    The relationships between the various concepts discussed in this article. Horizontal arrows from $A$ to $B$ signify that $B$ provides a context for understanding $A$, or equivalently, that $A$ pertains to a subset of $B$. For example, safe data and safe outputs are part of the Five Safes, and the Five Safes is a specialization of contextual integrity.
    Vertical arrows from $A$ to $B$ signify that $A$ is concerned with $B$. For example, a differential privacy definition is a criterion applied to SDC (statistical disclosure control) mechanisms, and `safe data' is concerned with the information type of a Type~\ref{eq:flow_data_researcher} flow.}
    \Description{}
    \label{fig:DP}
\end{figure*}

\section{Discussion}\label{secDiscussion}

\subsection{The wicked problem of disclosure control}

Disclosure control in statistical dissemination poses a ``wicked problem.'' Two defining characteristics of a wicked problem are 1) the relevant stakeholders have conflicting interests, and 2) the problem is so complex that there may not be consensus as to what the problem actually is, let alone a solution to it \citep{alford2017wicked}. %
In the absence of a framework such as CI or the FS, 
NSOs frequently frame SDC as a ``privacy--utility tradeoff.'' 
Although we regard this framing as problematic (see below), it serves to illustrate that SDC exhibits both characteristics of a wicked problem:
First, data custodians such as NSOs have a dual mandate to procure high quality data for its users while protecting the privacy of data subjects. The demand from data users (the \emph{recipients} of an information flow) for finer and more accurate information comes into direct conflict with the expectation of data \emph{subjects} for confidentiality and anonymity. Second, the fact that the privacy--utility tradeoff has remained as a loosely defined catchphrase demonstrates how successfully it has resisted attempts to be operationalized into a set of workable guidance, let alone one that commands wide agreement.

The FS makes a valuable contribution by articulating, and hence clarifying, an otherwise intractable problem. It reduces the ``wickedness'' of disclosure control decisions by modularizing a statistical dissemination into five manageable dimensions, along which an NSO may assess their risk tolerance and utility preference, subject to the pragmatic constraints they face. Moreover, %
the FS 
replaces the predominant ``privacy–utility tradeoff'' formulation of SDC with a more comprehensive representation of the problem and its possible solutions.
While the ``privacy versus utility'' framing may inject clarity and concreteness into an otherwise ill-posed problem, it is invariably reductive. It collapses the multidimensional nature of SDC into a tradeoff along a single axis. 
Further, this framing treats privacy simplistically as a matter of obscuring information or keeping it secret.
It is therefore important to recognize this framing for what it is: an ``\emph{obscurity}--utility tradeoff'' that fails to account for the full complexity of SDC. %
The FS, in contrast, articulates SDC in a way that brings clarity to many of its dimensions. %

However, it is one thing to properly articulate a problem, and a whole other to devise a solution. A data custodian looking to the FS and the FS only for an answer to their disclosure control question may be disappointed. As skeptics of the FS rightly point out \citep[see e.g.,][]{culnaneNotFitPurpose2020}, it only provides qualitative narratives; its five components do not come equipped with yardsticks that provide quantitative measures of safety. %

All is not lost in our view. The FS has demonstrated its value in instructing NSOs' decision-making by evaluating the risk of existing statistical dissemination paradigms (such as those reviewed in Section~\ref{sec:5s}) in each of its five dimensions. It is best viewed as a set of principles to reason with the potential factors that affect the disclosure risks of particular statistical dissemination choices. We may easily concede that these efforts in no way provide absolute measures of privacy for all conceivable dissemination choices, nor watertight guarantees of the absence of counterexamples. Such are the distinct advantages of technical criteria of privacy (e.g., DP) by virtue of their mathematical rigor. %
We believe, however, that while narrow conceptualizations of the disclosure control problem---such as technical privacy criteria---afford the apparent comfort of a concrete and precise answer, they sidestep the heart of the challenge: fostering a consensus about what the disclosure control problem really is. Indeed, as we explicated in Section~\ref{sec:DP}, criteria like DP pertain only to certain aspects of the `output' and `data' dimensions. That is to say, the precision and rigor of formal privacy notions invariably render them as insufficient answers to the broad, multidimensional problem of statistical dissemination. To this end, we reiterate our viewpoint that the FS, as a reparametrization of CI, provides a holistic construction of SDC, and hence a viable alternative angle for analysis.

\cut{\subsection{Future research}

Before closing, we briefly discuss four sets of important questions that remain unanswered in this work. %
First, our reparametrization of CI by the FS is faithful but inexact, in the sense that each framework captures some aspects of the disclosure control problem missing in the other. %
We aim to supplant the meanings that are lost in this translation in future research, in particular to detail aspects of `data' and `outputs' (from the FS) that go beyond the `subject(s)' and `information type' (from CI).%

Second, as we have already argued, DP offers a strong technical notion for assessing certain aspects of the `safe data' and `safe output' dimensions. Does there exist more comprehensive technical notions for these dimensions, and does there exist technical notions for the other dimensions of the FS? We surmise that a pursuit towards technicalization may not make sense universally. After all, some of the dimensions (such as safe people) may be too complex to allow mathematical tractability. However, others---such as safe settings---may benefit from advances in fields such as information security.

Third, we look to further explore the interaction between the FS and the legal frameworks governing the operations of the NSOs. %
For starters, we ask: what configurations of the FS best correspond to the current legal framework for a specific NSO? And how can the FS be used to update legal frameworks in the future?

Fourth, NSOs often require that the information they disseminate be used for ``statistical purposes'' \cite{hawesEvolutionInterpretationStatistical2026}. How should CI understand this requirement? %
Does it place a restriction on the `information type' parameter by requiring the flow to transmit population-level, as opposed to individual-level, information? Does it simultaneously imply some sort of transmission principle by mandating the information flow be for the public good?

}

\par\addvspace{1\baselineskip}
\noindent{\sffamily\textbf{Abbreviations:}} ABS, Australian Bureau of Statistics; API, application programming interface; CI, contextual integrity; DP, differential privacy; FS, Five Safes; FSRDC, Federal Statistical Research Data Center; GSF, Gold Standard File; IT, information technology; NA, not applicable; NSO, national statistical office; PUMS, Public Use Microdata Sample; SDS, Synthetic Data Server; SIPP, Survey of Income and Program Participation; SSB, SIPP Synthetic Beta; SDC, statistical disclosure control; UK, United Kingdom; US, United States; USCB, US Census Bureau.

\noindent{\sffamily\textbf{Acknowledgments:}} An earlier version of this paper was presented at the 5th Annual Symposium on Applications of Contextual Integrity on September 22, 2023 in Toronto, Canada. We thank the three anonymous symposium reviewers, other participants, John Eltinge, and Jeremy Seeman for helpful feedback on that version. 
Nevertheless, all errors and shortcomings in this work are solely attributable to ourselves. JB gratefully acknowledges partial financial support during this project from the Australian-American Fulbright Commission and the Kinghorn Foundation.

\noindent{\sffamily\textbf{Statement on conflicts of interest:}} The authors declare no conflicts of interest.

\noindent{\sffamily\textbf{Ethics statement:}} Ethical approval was not required for this article, as it did not involve research with human participants or the use of personal data.

\noindent{\sffamily\textbf{AI disclosure statement:}} The authors declare that no generative AI tools were used in the preparation of this manuscript.

\noindent{\sffamily\textbf{Data availability statement:}} No data were used or created in this work.

\bibliographystyle{ijpds-vancouver}
\bibliography{main}

\end{document}